\documentclass[pra,twocolumn,superscriptaddress,showpacs,preprintnumbers,amsmath,amssymb]{revtex4}

\usepackage{bm}
\usepackage{bbm}
\usepackage{amssymb}
\usepackage{amsfonts}
\usepackage{epsfig,graphicx}
\usepackage{amstext}
\usepackage{amsmath}
\usepackage{graphicx}
\usepackage{times}
\usepackage{txfonts}
\usepackage{dcolumn}
\usepackage{color}
\usepackage{dsfont}

\newcommand{\iden}{\openone}
\newcommand{\tr}{\mathrm{tr}}

\begin{document}

\title{Steered quantum coherence as a signature of quantum phase transitions in spin chains}

\author{Ming-Liang Hu}
\email{mingliang0301@163.com}
\affiliation{School of Science, Xi'an University of Posts and Telecommunications, Xi'an 710121, China}
\affiliation{Institute of Physics, Chinese Academy of Sciences, Beijing 100190, China}

\author{Yun-Yue Gao}
\affiliation{School of Science, Xi'an University of Posts and Telecommunications, Xi'an 710121, China}

\author{Heng Fan}
\email{hfan@iphy.ac.cn}
\affiliation{Institute of Physics, Chinese Academy of Sciences, Beijing 100190, China}
\affiliation{CAS Center for Excellence in Topological Quantum Computation, University of Chinese Academy of Sciences, Beijing 100190, China}
\affiliation{Songshan Lake Materials Laboratory, Dongguan 523808, China}

\begin{abstract}
We propose to use the steered quantum coherence (SQC) as a signature
of quantum phase transitions (QPTs). By considering various spin
chain models, including the transverse-field Ising model, \textit{XY}
model, and \textit{XX} model with three-spin interaction, we showed
that the SQC and its first-order derivative succeed in signaling
different critical points of QPTs. In particular, the SQC method is
effective for any spin pair chosen from the chain, and the strength
of SQC, in contrast to entanglement and quantum discord, is
insensitive to the distance (provided it is not very short) of the
tested spins, which makes it convenient for practical use as there
is no need for careful choice of two spins in the chain.
\end{abstract}


\maketitle

\section{Introduction}\label{sec:1}
Quantum coherence plays a fundamental role in the fields of quantum
optics \cite{Ficek} and thermodynamics \cite{ther5}. The resource
theoretic framework for quantifying coherence formulated in 2014
stimulates further study of it from a quantitative perspective
\cite{coher,Plenio,Hu}. In particular, it has been used to explain
the quantum advantage of many emerging quantum computation tasks,
including quantum state merging \cite{qsm}, deterministic quantum
computation with one qubit \cite{DQC1}, Deutsch-Jozsa algorithm
\cite{DJ}, and Grover search algorithm \cite{Grover}. The resource
theory of coherence also provides a basis for interpreting the wave
nature of a quantum system \cite{path1,path2} and the essence of
quantum correlations such as quantum entanglement \cite{coher-ent,
convex3,SQC,naqc2,naqc3,Tan} and various discordlike quantum
correlations \cite{DQC1,Tan,Yao,Hufan,Hux1,Yuc,Hux2}.

Besides the fundamental position in physics, quantum coherence is
also useful in studying critical behaviors of various spin chain
systems. For instance, the relative entropy of coherence for one
spin or two adjacent spins can detect quantum phase transitions
(QPTs) in the spin-1/2 transverse-field Ising, \textit{XX}, and
Kitaev honeycomb models \cite{chenj}, while critical behaviors of
the \textit{XY} model have been studied by virtue of the $l_1$ norm
of coherence \cite{Qin}. Moreover, the relative entropy and $l_1$
norm of coherence for two neighboring spins detect successfully the
Ising-type first-order QPT in the spin-1 \textit{XXZ} model
\cite{spin1}. The skew-information-based coherence measure
\cite{skif}, though it is not well defined \cite{Dubai}, can also
detect QPTs in certain spin chain models, including the spin-1/2
\textit{XY} model either without \cite{Karpat} or with three-spin
interaction \cite{Leisg,Liyc} and the spin-1/2 \textit{XYZ} model with
Dzyaloshinsky-Moriya interaction \cite{Ywl}.

In fact, other characterizations of quantumness in quantum
information science have also been used to study QPTs. One of them
is entanglement \cite{EoF}. Its role in exploring QPTs can be found
in Refs. \cite{nature,Osborne,Gusj1,Gusj2} and the review work
\cite{Amico}. Another quantumness measure is entropic quantum
discord \cite{QD,QD2}, which can detect QPTs in the \textit{XXZ}
model \cite{XXZ,Sarandy}, the transverse-field Ising model \cite{Ising,
Sarandy}, the transverse-field \textit{XY} model \cite{XY}, and the
\textit{XY} model with three-spin \cite{XYthree} or
Dzyaloshinsky-Moriya interaction \cite{XYDM}. Moreover, one can also
use geometric quantum discord to explore QPTs in certain spin chain
models \cite{Hu}. Nevertheless, although entanglement and quantum
discord were widely used to explore QPTs with great success,
entanglement is short ranged \cite{Amico}, so a careful choice of
two very short distance spins or the bipartition of the system is
required. Quantum discord, though can exist for two relatively
long-distance spins, its computation is NP complete \cite{qd-np}
(there is no closed formula even for a general two-qubit state
\cite{qdtwo}). These limit the scope of their applications in
exploring QPTs.

In this paper, we propose to use the steered quantum coherence (SQC)
\cite{SQC} as a signature of QPTs. We consider a general \textit{XY}
model with a transverse magnetic field and three-spin interaction, and
show that the SQC precisely signals all critical points of the QPTs.
In particular, compared with entanglement and quantum discord, the
SQC exists for any two spins in the chain, and its strength is
insensitive to the distance of two spins provided it is not very
short. This remarkable property of SQC releases the restriction on
the distance of the spin pair selected for probing QPTs and may
have important implications for experimental observation of QPTs as,
in general, it is hard to measure a weak quantity in experiments.
Moreover, different from quantum coherence of a state which is basis
dependent and one may extract useless information if the basis is
inappropriate, the SQC is analytically solvable for any two-spin
state and its value is definite. On the experimental side, the SQC
can be estimated by local projective measurements and one-qubit
tomography, which is also feasible with current techniques
\cite{qcexp1,qcexp2,qcexp3}. All the aspects above show that the SQC may
be a powerful tool to study QPTs in spin chain models.

The structure of this paper is as follows. In Sec. \ref{sec:2}, we
recall definition of the SQC and solution of the physical model.
Then in Sec. \ref{sec:3}, we discuss critical behaviors of SQC for
the considered model and show that it signals the QPTs precisely.
Finally, we summarize our main finding in Sec. \ref{sec:4}.

\section{Preliminaries}\label{sec:2}
We first present definition of the SQC. For a state $\rho_{AB}$ with
the two qubits held, respectively, by Alice and Bob, the SQC was
defined by Alice's local measurements and classical communication
between Alice and Bob. To be explicit, Alice carries out one of the
pre-agreed measurements $\{\sigma^\mu\}_{\mu=x,y,z}$ ($\sigma^\mu$
is the Pauli operator) on qubit $A$ and communicates to Bob her
choice $\sigma^\mu$. Then Bob's system collapses to the
ensemble states $\{p_{\mu,a}, \rho_{B|\Pi_\mu^a}\}$, with $p_{\mu,a}
=\tr(\Pi_\mu^a \rho_{AB})$ being the probability of Alice's outcome
$a\in\{0,1\}$, and $\rho_{B|\Pi_\mu^a}= \tr_A (\Pi_\mu^a\rho_{AB})
/p_{\mu,a}$ being Bob's conditional state. Moreover, $\Pi_\mu^{a}=
[\openone_2+(-1)^a \sigma^\mu]/2$ is the measurement operator and
$\iden_2$ is the identity operator.

For Alice's chosen observable $\sigma^\mu$, Bob can measure the
coherence of the ensemble $\{p_{\mu,a}, \rho_{B|\Pi_\mu^a}\}$ with
respect to the eigenbasis of either one of the remaining two Pauli
operators. After Alice's all possible measurements
$\{\Pi_\mu^a\}_{\mu=x,y,z}$ with equal probability, the SQC at Bob's
hand can be defined as the following averaged quantum coherence
\cite{SQC}
\begin{equation} \label{eq2a-2}
 C^{na}(\rho_{AB})= \frac{1}{2}\sum_{\mu,\nu,a\atop \mu\neq\nu}
                    p_{\mu,a} C^{\sigma^\nu}(\rho_{B|\Pi_\mu^a}),
\end{equation}
where $C^{\sigma^\nu}(\rho_{B|\Pi_\mu^a})$ is the coherence of
$\rho_{B|\Pi_\mu^a}$ defined in the reference basis spanned by the
eigenbases of $\sigma^\nu$ \cite{coher}.

In this paper, we use the $l_1$ norm of coherence and the relative
entropy of coherence which are favored for their ease of
calculation. By denoting $\{|\psi_i\rangle\}$ the eigenbases of
$\sigma^\nu$, their analytical solutions are given, respectively, by
\cite{coher}
\begin{equation} \label{eq2a-3}
 \begin{split}
  & C_{l_1}^{\sigma^\nu}(\rho)= \sum_{i\neq j}|\langle\psi_i|\rho|\psi_j\rangle|,\\
  & C_{re}^{\sigma^\nu}(\rho)= -\sum_i \langle\psi_i|\rho|\psi_i\rangle
                               \log_2 \langle\psi_i|\rho|\psi_i\rangle-S(\rho),
 \end{split}
\end{equation}
with $S(\rho)=-\tr (\rho\log_2 \rho)$ denoting the von Neumann
entropy. Based on these formulas, one can then obtain the
corresponding SQC $C_{l_1}^{na}(\rho_{AB})$ and $C_{re}^{na}
(\rho_{AB})$.

Next, we introduce the \textit{XY} model with a transverse magnetic
field and three-spin interaction. The Hamiltonian for such a model
can be written as
\begin{equation} \label{eq2b-1}
 \begin{split}
  \hat{H}= &-\sum\limits_{n=1}^N\left(\frac{1+\gamma}{2}{\sigma_n^x\sigma_{n+1}^x}
            +\frac{1-\gamma}{2}{\sigma_n^y\sigma_{n+1}^y}+\lambda\sigma_n^z\right) \\
           &-\sum\limits_{n=1}^N\alpha({\sigma_{n-1}^x\sigma_n^z\sigma_{n+1}^x}
            +{\sigma_{n-1}^y\sigma_n^z\sigma_{n+1}^y}),
 \end{split}
\end{equation}
where $\sigma_n^\mu$ ($\mu=x,y,z$) are the Pauli operators at site
$n$, $\lambda$ is the transverse magnetic field, $\gamma$ denotes the
anisotropy of the system arising from the nearest-neighbor
interaction, and $\alpha$ denotes the strength of the three-spin
interaction arising from the next-to-nearest-neighbor interaction
\cite{three}. Moreover, $N$ is the number of spins in the chain, and
we assume the periodic boundary conditions.

The Hamiltonian $\hat{H}$ can be diagonalized by first using the
Jordan-Wigner transformation \cite{QPTs2}
\begin{equation} \label{eq2b-2}
 \begin{split}
  & \sigma_n^x=\prod_{m<n}\left(1-2{c_m^\dag}c_m\right)\left(c_n+c_n^\dag\right), \\
  & \sigma_n^y=-i\prod_{m<n}\left(1-2{c_m^\dag}c_m\right)\left(c_n-c_n^\dag\right),~
    \sigma_n^z=1-2 c_n^\dag c_n,
 \end{split}
\end{equation}
which maps the spins to spinless fermions with the creation
(annihilation) operators $c_n^\dag$ ($c_n$). Then by virtue of the
Fourier transformation $\tilde{c}_k=\sum_l c_l e^{-ilx_k}/\sqrt{N}$
($x_k=2\pi k/N$) and the Bogoliubov transformation $d_k =
\cos(\theta_k/2)\tilde{c}_k-i\sin(\theta_k/2)\tilde{c}_{-k}^\dag$,
one can obtain \cite{epjb}
\begin{equation} \label{eq2b-3}
 \hat{H}=\sum_{k=-M}^M 2\varepsilon_k\left(d_k^\dag d_k-\frac{1}{2}\right),
\end{equation}
where $M=(N-1)/2$, $\theta_k=\arcsin[-\gamma \sin
(x_k)/\varepsilon_k]$, and the energy spectrum is given by
\begin{equation} \label{eq2b-4}
 \varepsilon_k=\sqrt{\epsilon_k^2+ \gamma^2\sin^2(x_k)},
\end{equation}
with $\epsilon_k= \lambda-\cos(x_k)-2\alpha\cos(2x_k)$.

To calculate the SQC, one needs to obtain the density operator
$\rho_{i,i+r}$ for the spin pair $(i,i+r)$, with $r$ denoting the
distance of two spins in units of the lattice constant. In the Bloch
representation, $\rho_{i,i+r}$ can always be decomposed as
\begin{equation} \label{eq2b-5}
 \rho_{i,i+r}=\frac{1}{4}\sum_{\mu,\nu} t_{\mu\nu}\sigma_i^\mu\otimes \sigma_{i+r}^\nu,
\end{equation}
where $\mu,\nu\in \{0,x,y,z\}$, $t_{\mu\nu}= \tr (\rho_{i,i+r}
\sigma_i^\mu\otimes \sigma_{i+r}^\nu)$, and $\sigma_i^0=\iden_2$.
Due to the translation invariance, $\rho_{i,i+r}$ will be
independent of the position $i$ and depends only on the distance $r$
of two spins. Then one can obtain the nonzero $t_{\mu\nu}$ of
$\rho_{i,i+r}$ as \cite{Wang,Gusj}
\begin{equation} \label{eq2b-6}
 t_{z0}=t_{0z}=\langle\sigma^z\rangle,~
 t_{\mu\mu}=\langle\sigma_{i}^\mu\sigma_{i+r}^\mu\rangle~(\mu\in\{x,y,z\}),
\end{equation}
where $\langle\sigma^z\rangle$ is the magnetization intensity given
by \cite{magnet}
\begin{equation} \label{eq2b-7}
 \langle\sigma^z\rangle= \frac{1}{N} \sum_k
                         \frac{\epsilon_k\tanh(\beta\varepsilon_k)}{\varepsilon_k},
\end{equation}
and $\beta=1/k_B T$, with $k_B$ being the Boltzmann constant.
Moreover, the spin-spin correlation functions are given by
\cite{xyt1}
\begin{equation} \label{eq2b-8}
 \begin{split}
  \langle\sigma_i^x\sigma_{i+r}^x\rangle&=
   \begin{vmatrix}
    G_{-1}  & G_{-2}  & \cdots & G_{-r}   \\
    G_{0}   & G_{-1}  & \cdots & G_{-r+1} \\
    \vdots  & \vdots  & \ddots & \vdots   \\
    G_{r-2} & G_{r-3} & \cdots & G_{-1}
   \end{vmatrix},\\
  \langle\sigma_i^y\sigma_{i+r}^y\rangle&=
   \begin{vmatrix}
    G_{1}   & G_{0}   & \cdots & G_{-r+2} \\
    G_{2}   & G_{1}   & \cdots & G_{-r+3} \\
    \vdots  & \vdots  & \ddots & \vdots   \\
    G_{r}   & G_{r-1} & \cdots & G_{1}
   \end{vmatrix},
 \end{split}
\end{equation}
and $\langle\sigma_i^z\sigma_{i+r}^z\rangle= {\langle\sigma_i^z
\rangle}^2 -G_r G_{-r}$, where $G_n$ ($-r\leqslant n\leqslant r$) is given by
\begin{equation} \label{eq2b-9}
 G_n= -\sum_k\frac{[\cos(n x_k)\epsilon_k+\gamma\sin(n x_k)\sin(x_k)]
      \tanh\left(\beta\varepsilon_k\right)}{N\varepsilon_k}.
\end{equation}

For the two-spin density operator $\rho_{i,i+r}$ with its nonzero
elements constrained by Eq. \eqref{eq2b-6}, the SQC can be obtained
analytically as
\begin{eqnarray} \label{eq2b-10}
 \begin{aligned}
  C_{l_1}^{na}(\rho_{i,i+r})= & t_{0z}+\frac{1}{2}\left(t_{xx}+t_{yy}+\sqrt{t_{0z}^2+t_{xx}^2}
                                +\sqrt{t_{0z}^2+t_{yy}^2}\right), \\
   C_{re}^{na}(\rho_{i,i+r})= & 2-H_2(\tau_1)-H_2(\tau_2)-\frac{(1+t_{z0})H_2(\tau_3)}{2} \\
                              & -\frac{(1-t_{z0})H_2(\tau_4)}{2}+H_2\left(\frac{1+t_{0z}}{2}\right),
 \end{aligned}
\end{eqnarray}
where $H_2(\cdot)$ denotes the binary Shannon entropy function, and
the parameters $\tau_i$ ($i=1,2,3,4$) are given by
\begin{equation} \label{eq2b-11}
 \begin{aligned}
  & \tau_1=\frac{1}{2}\left(1 + \sqrt{t_{0z}^2+t_{xx}^2}\right),~
    \tau_2=\frac{1}{2}\left(1 + \sqrt{t_{0z}^2+t_{yy}^2}\right), \\
  & \tau_3=\frac{1}{2}+ \frac{|t_{0z}+t_{zz}|}{2(1+t_{z0})},~
    \tau_4=\frac{1}{2}+ \frac{|t_{0z}-t_{zz}|}{2(1-t_{z0})}.
 \end{aligned}
\end{equation}

\section{SQC and QPTs in spin chain models}\label{sec:3}
Based on the above preliminaries, we discuss in this section
critical behaviors of the spin chain described by Eq. \eqref{eq2b-1}
by using the SQC. We show that the extreme points of the SQC for
any two spins as well as the discontinuity of its first derivative
are able to indicate QPTs in the considered model.

\subsection{Transverse-field Ising model}\label{sec:3a}
To begin with, we consider the transverse-field Ising model which
corresponds to $\gamma=1$ and $\alpha=0$ in Eq. \eqref{eq2b-1}. For
such a model, it is known that there is a second-order QPT at
$\lambda_c=1$. At this point, the global phase flip symmetry breaks
and the correlation length diverges \cite{Amico}.

\begin{figure}
\centering
\resizebox{0.41 \textwidth}{!}{%
\includegraphics{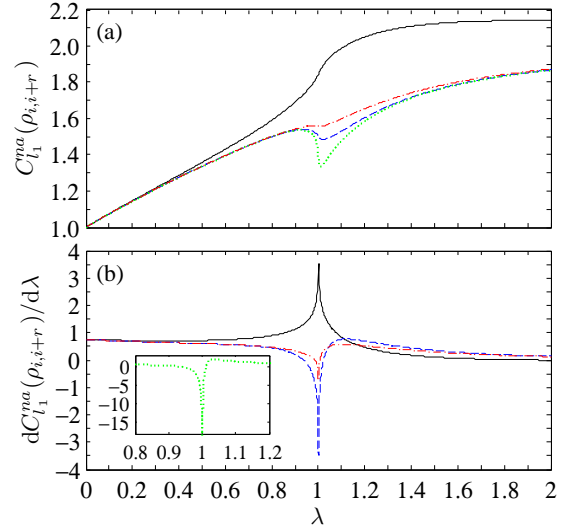}}
\caption{$C_{l_1}^{na}(\rho_{i,i+r})$ (a) and its first derivative
$\mathrm{d}C_{l_1}^{na}(\rho_{i,i+r})/\mathrm{d}\lambda$ (b) versus
$\lambda$ for the Ising model with $N=2001$. The solid black,
dash-dotted red, dashed blue, and dotted green lines correspond to
$r=1$, 5, 10, and 100, respectively. The dotted green line
in panel (b) is shown in the inset to better visual the QPT.} \label{fig:1}
\end{figure}

To reveal that the SQC can indicate QPTs in the Ising model, we
show in Fig. \ref{fig:1} the dependence of $C_{l_1}^{na}
(\rho_{i,i+r})$ and its first derivative on $\lambda$ with different
distances $r$ of the spin pair. For $r\leqslant 3$, $C_{l_1}^{na}
(\rho_{i,i+r})$ increases monotonically with the increase of
$\lambda$, and its first-order derivative with respect to $\lambda$
shows a discontinuity at $\lambda_c=1$. For the tested spins with long
distances ($r\geqslant 4$), as depicted in Fig. \ref{fig:1}(a),
$C_{l_1}^{na} (\rho_{i,i+r})$ does not behave as a monotonic
increasing function of $\lambda$. Instead, there exists a pronounced
cusp close to $\lambda_c=1$. A further numerical calculation shows
that the critical point $\lambda_t$ for the minimum of this cusp
approaches monotonically to $\lambda_c$ with the increase of $r$,
e.g., $\lambda_t-\lambda_c \sim 10^{-6}$ when $r=1000$ and $N=2001$.
Then it is reasonable to conclude that for an infinite chain, the
minimum of this cusp can precisely signal the QPT at $\lambda_c=1$
when $r$ is very large. Moreover, one can observe from Fig.
\ref{fig:1}(b) that the discontinuity of $\mathrm{d}C_{l_1}^{na}
(\rho_{i,i+r})/ \mathrm{d}\lambda$ indicates the QPT at
$\lambda_c=1$ for the chosen tested spins with any distance.

\begin{figure}
\centering
\resizebox{0.41 \textwidth}{!}{%
\includegraphics{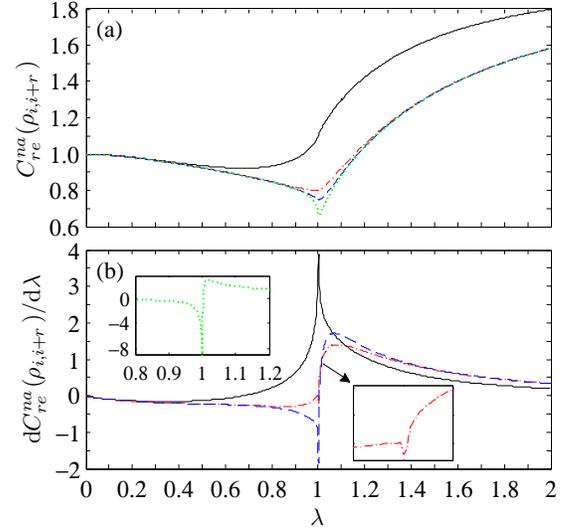}}
\caption{$C_{re}^{na}(\rho_{i,i+r})$ (a) and its first derivative
$\mathrm{d}C_{re}^{na}(\rho_{i,i+r})/\mathrm{d}\lambda$ (b) versus
$\lambda$ for the Ising model with $N=2001$. The solid black,
dash-dotted red, dashed blue, and dotted green lines correspond to
$r=1$, 5, 10, and 100, respectively. The inset in the bottom
right corner is an amplified plot of the dash-dotted red line
in the neighborhood of $\lambda_c$, and the dotted green line
in panel (b) is shown in the top left corner to better visual the
QPT.} \label{fig:2}
\end{figure}

With the same system parameters as in Fig. \ref{fig:1}, we displayed
in Fig. \ref{fig:2} dependence of $C_{re}^{na}(\rho_{i,i+r})$ and
its first derivative on $\lambda$. One can see that with the
increasing strength of the transverse magnetic field $\lambda$,
$C_{re}^{na} (\rho_{i,i+r})$ first decreases to a minimum, and then
turns to be increased gradually. As for $\rho_{i,i+r}$ with large
$r$, $C_{re}^{na}(\rho_{i,i+r})$ also shows a pronounced cusp in the
neighborhood of $\lambda_c$, and with the increase of $r$, the
critical point of $\lambda_t$ for the minimum of this cusp
approaches to $\lambda_c$ more rapidly than that for $C_{l_1}^{na}
(\rho_{i,i+r})$, e.g., $\lambda_{t}-\lambda_c \sim 10^{-8}$ for
$r=1000$ and $N=2001$. This suggests that the cusp of SQC can signal
the QPT taking place at $\lambda_c$ for two long-distance tested
spins. Moreover, the first-order derivative of $C_{re}^{na}
(\rho_{i,i+r})$, as expected, also presents a discontinuity at the
phase transition point $\lambda_c=1$ for two spins with different
distances.

All the above observations show evidently that the SQC and its first-order
derivative for any two spins can clearly indicate QPT in the
Ising model. In particular, one can see from Figs. \ref{fig:1} and
\ref{fig:2} that beyond the adjacent region of $\lambda_c$, the
curves of SQC for two spins with different large $r$ are nearly
overlapped; i.e., there is almost no decrease of the SQC for
$\rho_{i,i+r}$ with different large $r$. Such a property can be
immediately applied to reduce the experimental demands to detect
QPTs, as one can choose two spins at any distance to achieve the
same feat.

We have also checked efficiency of other signatures of QPT. For
entanglement and quantum discord, the discontinuities of their first
derivatives can detect QPTs in the Ising chain \cite{Ising}. But the
entanglement exists only for $r\leqslant 2$, and hence imposes a strict
restriction on the distance of the tested spins, while the calculation
of quantum discord is a hard task even when $\rho_{i,i+r}$ is
available \cite{qdtwo}. Moreover, it can be seen from Eqs.
\eqref{eq2b-5} and \eqref{eq2b-6} that the one-spin coherence is
always zero. As for the two-spin coherence, its derivative shows a
discontinuity at $\lambda_c$, but its estimation needs a two-qubit
state tomography.

\subsection{Transverse-field \textit{XY} model}\label{sec:3b}
Next, we consider the transverse-field \textit{XY} model, which
corresponds to $\alpha=0$ in Eq. \eqref{eq2b-1}. There are two QPTs
\cite{phase,QPTxy}. The first one occurs at $\lambda_c= 1$. For
$\lambda< \lambda_c$, the system is in the ferromagnetic ordered
phase, while for $\lambda> \lambda_c$ it is in the paramagnetic
quantum disordered phase. The second one occurs at $\gamma_c=0$ and
$\lambda\in (0,1)$. It further separates the ferromagnetic ordered
phase into two regions, i.e., the ferromagnet ordered along either
the $x$ ($\gamma<0$) or the $y$ ($\gamma>0$) axis.

\begin{figure}
\centering
\resizebox{0.41 \textwidth}{!}{%
\includegraphics{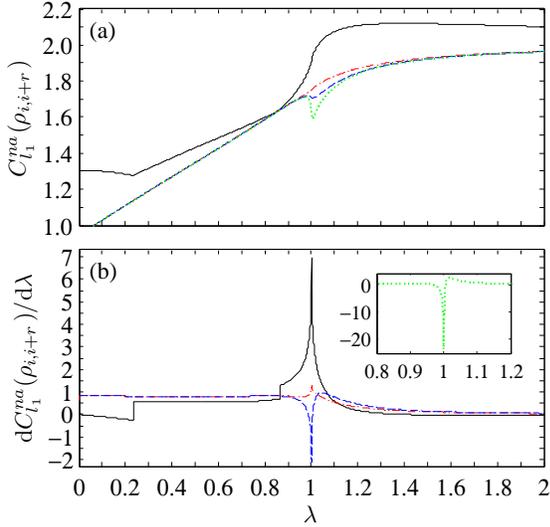}}
\caption{$C_{l_1}^{na}(\rho_{i,i+r})$ (a) and its first derivative
$\mathrm{d}C_{l_1}^{na}(\rho_{i,i+r})/ \mathrm{d} \lambda$ (b)
versus $\lambda$ for the \textit{XY} model with $\gamma=0.5$ and
$N=2001$. The solid black, dash-dotted red, dashed blue, and
dotted green lines correspond to $r=1$, 5, 10, and 100,
respectively. The dotted green line in panel (b) is shown in the
inset to better visual the QPT.} \label{fig:3}
\end{figure}

In Fig. \ref{fig:3}, we show the dependence of $C_{l_1}^{na}
(\rho_{i,i+r})$ and its first derivative on $\lambda$ for the
\textit{XY} model with $\gamma=0.5$. For two neighboring spins, the
discontinuity of $\mathrm{d}C_{l_1}^{na}(\rho_{i,i+r})/ \mathrm{d}
\lambda$ precisely signals the QPT at $\lambda_c$, and there exist
two inflexions for it, which are not critical points of QPTs
\cite{xyt1,xyt2}. When $r$ is large, the curves of
$C_{l_1}^{na}(\rho_{i,i+r})$ with different $r$ are nearly
overlapped beyond the adjacent region of $\lambda_c$, and there
exists an abrupt cusp in the neighborhood of $\lambda_c$. The
critical point of $\lambda_t$ corresponds to the minimum of this cusp
approaches asymptotically to $\lambda_c$ with the increase of $r$,
e.g., $\lambda_t-\lambda_c \sim 10^{-7}$ when $r=1000$ and $N=2001$.
Similar to the Ising model, the insensitivity of the SQC to the distance
(provided it is not very short) of the tested spins in the \textit{XY}
chain also has important practical consequences for experimental
characterization of QPTs. With regard to the first-order derivative of
$C_{l_1}^{na}(\rho_{i,i+r})$, it shows a discontinuity at
$\lambda_c$, irrespective of $r$. Hence, it is able to precisely detect
the QPT for two spins at any distance.

\begin{figure}
\centering
\resizebox{0.41 \textwidth}{!}{%
\includegraphics{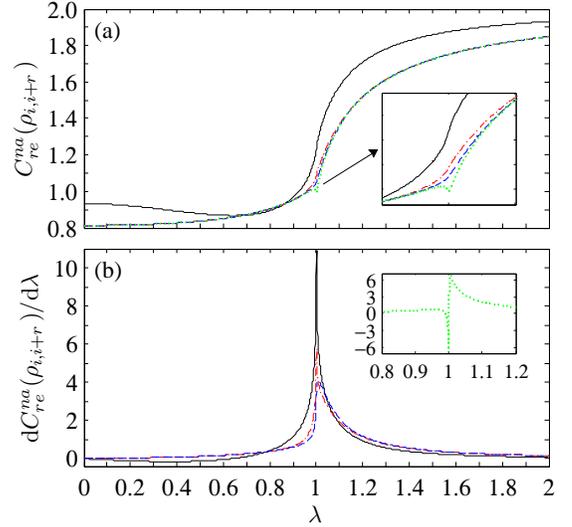}}
\caption{$C_{re}^{na}(\rho_{i,i+r})$ (a) and its first derivative
$\mathrm{d}C_{re}^{na}(\rho_{i,i+r})/ \mathrm{d}\lambda$ (b) versus
$\lambda$ for the \textit{XY} model with $\gamma=0.5$ and $N=2001$.
The solid black, dash-dotted red, dashed blue, and dotted green
lines correspond to $r=1$, 5, 10, and 100, respectively. The
inset in panel (a) is an amplified plot of the lines in the neighborhood
of $\lambda_c$, and the dotted green line in panel (b) is shown in the
inset to better visual the QPT.} \label{fig:4}
\end{figure}

Similarly, we show in Fig. \ref{fig:4} the capability of $C_{re}^{na}
(\rho_{i,i+r})$ and its derivative in detecting QPT at $\lambda_c=1$.
First, for two spins with long distances, the curves of
$C_{re}^{na} (\rho_{i,i+r})$ are nearly overlapped for $\lambda$
deviating from the adjacent region of $\lambda_c$. On the contrary,
there is a cusp close to $\lambda_c$, and the critical $\lambda_t$
related to the bottom of this cusp approaches rapidly to $\lambda_c$
with the increase of $r$, e.g., $\lambda_t-\lambda_c\sim 10^{-10}$
when $r=1000$ and $N=2001$. Second, the first derivative of
$C_{re}^{na} (\rho_{i,i+r})$ shows a discontinuity at $\lambda_c$,
irrespective of the distance of the spin pair in the chain. This
indicates that the phase transition point in the \textit{XY} model
can also be signaled precisely by $\mathrm{d}C_{re}^{na}
(\rho_{i,i+r})/ \mathrm{d}\lambda$.

We have also examined QPTs of the \textit{XY} model at $\gamma_c=0$
and $\lambda\in (0,1)$. For conciseness of this paper, we do not
present the plots here. The numerical calculation shows that this
QPT can be signaled precisely by the extremal behaviors of the SQC.
To be explicit, $C_{l_1}^{na}(\rho_{i,i+r})$ is maximal for $r=1$
and minimal for $r\geqslant 2$ at $\gamma_c$, while $C_{re}^{na}
(\rho_{i,i+r})$ always reaches to its minimum at $\gamma_c$.
However, there is no extremal, discontinuous, or singular behavior
being observed for the first-order derivative of the SQC with respect to
the anisotropic parameter $\gamma$.

As for concurrence of $\rho_{i,i+r}$, it is non-null for two spins
with very short distance; e.g., for $\gamma= 0.5$, its first
derivative detects the QPT at $\lambda_c$ only when $r\leqslant 3$.
The critical point $\lambda_c$ can also be detected by the first
derivative of quantum discord for two spins more distant than second
neighbors \cite{XY}, and similarly for the two-spin coherence.
However, the strength of quantum discord and two-spin coherence
decrease as we increase $r$, especially in the region of $\lambda>
\lambda_c$, hence it is hard to detect them experimentally when $r$
is large.

\subsection{Transverse-field \textit{XX} model with three-spin interaction}\label{sec:3c}
Now, we consider a more general case where only $\gamma=0$ is
assumed in Eq. \eqref{eq2b-1}. The ground-state phase diagram
consists of four sectors \cite{epjb}: the spin-saturated phase in
the regions of $\lambda>\lambda_{c_1}$ and $\lambda<\lambda_{c_i}$
($i= 2$ when $\alpha<1/8$ and $i=3$ otherwise), the spin liquid
\uppercase\expandafter{\romannumeral 1} phase in the region of
$\lambda\in(\lambda_{c_2},\lambda_{c_1})$, and the spin liquid
\uppercase\expandafter{\romannumeral 2} phase in the region of
$\lambda\in(\lambda_{c_3},\lambda_{c_2})$ and $\alpha>1/8$. Here,
$\lambda_{c_1,c_2}= 2\alpha\pm 1$ and $\lambda_{c_3}=
-(1+32\alpha^2)/16\alpha$.

\begin{figure}
\centering
\resizebox{0.42 \textwidth}{!}{%
\includegraphics{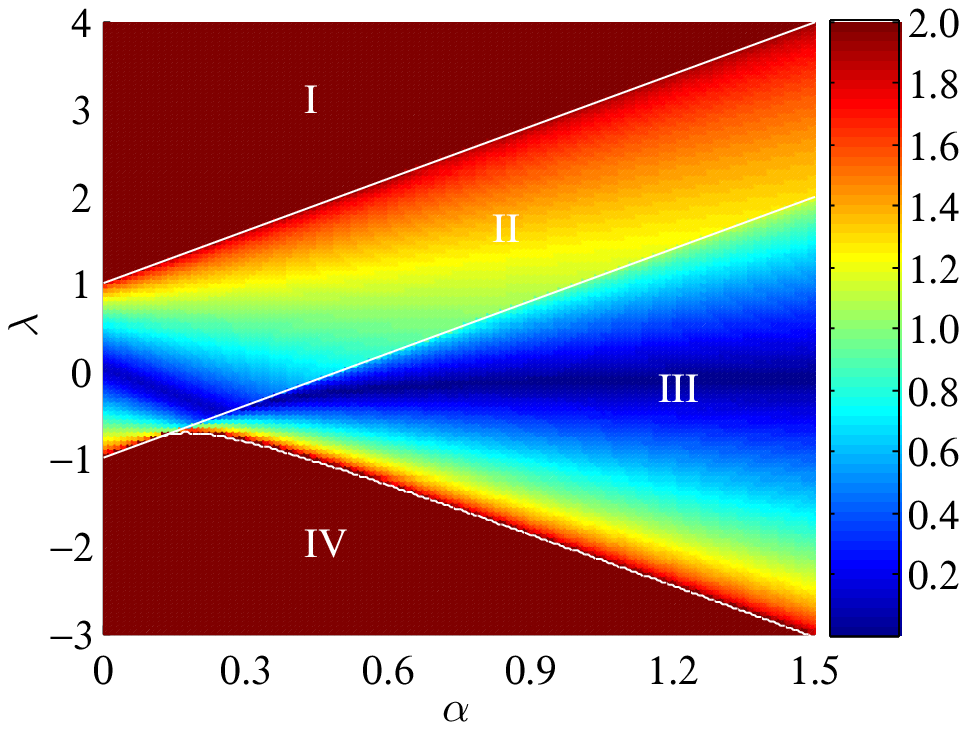}}
\centering
\resizebox{0.42 \textwidth}{!}{%
\includegraphics{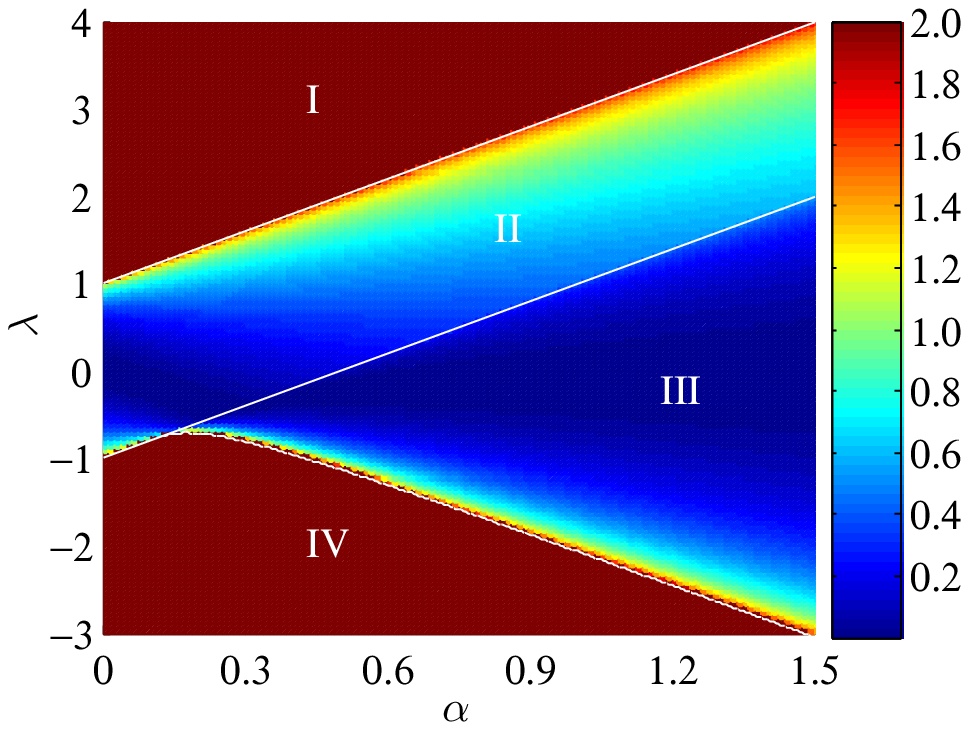}}
\caption{$C_{l_1}^{na}(\rho_{i,i+r})$ (top) and $C_{re}^{na}
(\rho_{i,i+r})$ (bottom) versus $\alpha$ and $\lambda$ for the
three-spin interaction \textit{XX} model with $N=2001$ and $r=100$.
Here, regions I and IV correspond to spin-saturated phase, while
regions II and III correspond to two kinds of spin liquid phases.}
\label{fig:5}
\end{figure}

In Fig. \ref{fig:5}, we plot the SQC as functions of $\alpha$ and
$\lambda$ for the three-spin interaction \textit{XX} model with
$N=2001$ and $r=100$. As can be seen from this figure, both
$C_{l_1}^{na}(\rho_{i,i+r})$ and $C_{re}^{na}(\rho_{i,i+r})$ can
signal the regions of different phases. To be explicit, when the
system is in the spin-saturated phase, the two SQC measures take
their values of about 2, while in the two spin liquid phases, one
can observe a pronounced decrease of their values. The critical
lines (i.e., $\lambda= \lambda_{c_1}$ and $\lambda= \lambda_{c_3}$)
separating the spin-saturated phase from the spin liquid phase
correspond to two inflexions of the SQC. For $\alpha> 1/8$, the
boundary (i.e., $\lambda=\lambda_{c_2}$) between the spin liquid I
and spin liquid II phases corresponds to another inflexion of the
SQC. Besides the three critical lines, there is a critical line
indicated by the minimum of the SQC, but as was shown in Ref.
\cite{epjb}, it is not a boundary of QPT.

\begin{figure}
\centering
\resizebox{0.42 \textwidth}{!}{%
\includegraphics{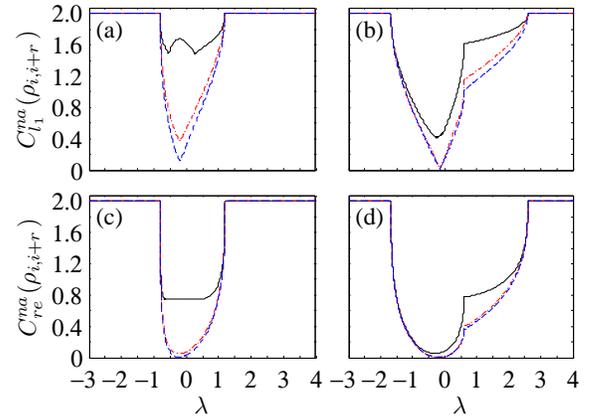}}
\caption{$C_{l_1}^{na}(\rho_{i,i+r})$ [panels (a) and (b)] and
$C_{re}^{na}(\rho_{i,i+r})$ [panels (c) and (d)] versus $\lambda$
for the three-spin interaction \textit{XX} model with $N=2001$.
Here, $\alpha=0.1$ for panels (a) and (c), $\alpha=0.8$ for panels (b) and (d).
The solid black, dash-dotted red, and dashed blue lines (from
top to bottom) correspond to $r=1$, 10, and 100, respectively.} \label{fig:6}
\end{figure}

To gain more insight into the critical behaviors of SQC for the
present model, we further plot in Fig. \ref{fig:6} the dependence of
$C_{l_1}^{na}(\rho_{i,i+r})$ and $C_{re}^{na} (\rho_{i,i+r})$ on
$\lambda$ with different $\alpha$ and $r$. Besides those behaviors
observed in Fig. \ref{fig:5}, one can observe that when $r=1$ and
$\alpha< 1/8$, there are two cusplike minima which are pronounced
for $C_{l_1}^{na}(\rho_{i,i+r})$ and are not obvious for
$C_{re}^{na} (\rho_{i,i+r})$, but they are not critical points of
QPTs \cite{epjb}. In this sense, the SQCs of long-distance spin
pairs are more reliable than that of the neighboring spin pair in
detecting QPTs of the three-spin interaction \textit{XX} model.
Looking at Fig. \ref{fig:6}, one can note that the curves of SQC for
the spin pairs with different long distances are nearly overlapped;
that is, the SQC in this model is also insensitive to the variation
of the distance (provided it is not very short) of two spins. Such a
property will be useful in the experimental detection of QPTs where
other characterizations of quantumness are very weak and hence
cannot be detected efficiently.

As for concurrence of $\rho_{i,i+r}$, it is able to detect partial
QPTs in the three-spin interaction model for the spin pair
with small $r$ \cite{XYthree}. But when $r$ is large, its value
becomes very small, and the regions of non-null concurrence shrink
to the vicinity of $\lambda_{c_2}$ (if $\alpha< 1/8$) or
$\lambda_{c_3}$ (if $\alpha> 1/8$). The quantum discord is a
reliable indicator of QPTs when choosing two neighboring spins
\cite{XYthree}, and the two-spin coherence can detect the QPTs as
well for small $r$. However, they also decrease with an increase in
$r$, especially when $\alpha> 1/8$ and $r$ is large, they both
oscillate rapidly with respect to $\lambda$ in the region of
$\lambda\in (\lambda_{c_3}, \lambda_{c_2})$, with a large number of
extreme points being observed. It is therefore hard to distinguish
these points from the critical points of QPTs.

Finally, we present an explanation for the underpinning of the
observed phenomena in the above subsections, that is, the
insensitivity of the SQC to the distance $r$ of two spins in the chain
and the divergence in the derivative of the SQC with respect to the
magnetic field $\lambda$. For brevity, we consider the Hamiltonian
$\hat{H}$ without the three-spin interaction, and the general
$\hat{H}$ of Eq. \eqref{eq2b-1} can be analyzed in a similar manner.

First, we explain the insensitivity of the SQC to $r$. As $t_{0z}$
is independent of $r$, one only needs to consider the $r$ dependence
of $t_{\mu\mu}$ which are determined by $\{G_n\}_{n=-r}^{r}$. From
Eq. \eqref{eq2b-9}, one can obtain that for $\gamma=0$, $|G_{\pm 1}|$
is maximal among all $\{|G_n|\}$ if $\lambda \lesssim 0.6736$ and
$|G_{0}|$ is maximal if $\lambda\gtrsim 0.6736$, while for
$\gamma\in(0,1]$, $|G_{-1}|$ is maximal if $\lambda<\lambda_{0}$ and
$|G_{0}|$ is maximal if $\lambda> \lambda_{0}$, with $\lambda_{0}$
increasing from 0.6736 to 1 when $\gamma$ increases from 0 to 1.
Moreover, $|G_{\pm n}|$ with large $n$ are negligible compared with
those with small $n$. For example, for the Ising model, we have
$G_n=-2/[(2n+1)\pi]$ at $\lambda=\lambda_c$, $G_{-1}=1$ and
$G_{n}=0$ ($n\neq -1$) at $\lambda=0$ in the thermodynamic limit
($N\rightarrow \infty$), while for the \textit{XX} model, we have
$G_0=2\theta_0/\pi-1$ and $G_n=2\sin(n\theta_0)/(n\pi)$ ($n\neq 0$),
where $\theta_0=\arccos(\min\{\lambda,1\})$. Therefore, for the
Ising model, $|G_n/G_{-1}|=1/(2n+1)$ at $\lambda=\lambda_c$, and
such a ratio will be further decreased when $\lambda$ deviates from
$\lambda_c$. Similarly, for the \textit{XX} model, $|G_n/G_{\pm 1}|=
|\sin(n\theta_0)|/(n\sin\theta_0)$ and $|G_n/G_0|=|\sin(n\theta_0)|/
[n(\pi -2\theta_0)]$. As a consequence, even when $r$ is very large,
only those terms $G_{\pm n}$ with small $n$ dominate in $t_{xx}$ and
$t_{yy}$, and this results in the insensitivity of $C_{l_1}^{na}
(\rho_{i,i+r})$ to large $r$. Moreover, it is easy to see that
$t_{zz}$ depends weakly on large $r$, thus $C_{re}^{na}(\rho_{i,i+r})$
is also insensitive to large $r$.

Physically, the insensitivity of the SQC indicator to the distance
between the tested spins can also be comprehended from the fact that
the SQC is null only for $\rho_{AB}= \rho_A \otimes \openone_2/2$ as
it takes into account the three mutually unbiased bases \cite{SQC}.
That is, it characterizes a more general form of correlation and
could exist in a parameter region in which there are no
entanglement and quantum discord. In fact, the insensitivity of the
SQC indicator to large $r$ also has its roots in the insensitivity
of the elements of the reduced density matrices $\rho_{i,i+r}$ with
large $r$. But for these $\rho_{i,i+r}$, the entanglement has
already disappeared and the quantum discord is very weak.
Moreover, some sudden change points of quantum discord may not
correspond to QPTs as they are caused by the optimization procedure
in its definition \cite{Karpat}.

Second, we explain the divergence in the derivative of the SQC
with respect to $\lambda$. Given that $T=0$, then from Eqs.
\eqref{eq2b-7} and \eqref{eq2b-9} one can obtain
\begin{equation} \label{eq3-1}
 \begin{aligned}
  & \frac{\partial{t_{0z}}}{\partial{\lambda}}= \frac{\gamma^2}{N}\sum_k \frac{\sin^2(x_k)}{\varepsilon_k^3}, \\
  & \frac{\partial{G_n}}{\partial{\lambda}}= \frac{\gamma}{N}\sum_k\frac{\epsilon_k \sin(nx_k)\sin(x_k)
                                             -\gamma\cos(nx_k)\sin^2(x_k)}{\varepsilon_k^3},
 \end{aligned}
\end{equation}
from which one can see that both $\partial{t_{0z}}/\partial{\lambda}$
and $\partial{G_n}/\partial{\lambda}$ are divergent at $\lambda=
\lambda_c$ as the two fractions in the above equation approach
infinity. For the \textit{XX} model, one can see more specifically the
divergence of $\partial{t_{0z}}/\partial{\lambda}$ and $\partial{G_n}/
\partial{\lambda}$. This is because in the thermodynamic limit, we have
$\partial{t_{0z}}/\partial{\lambda}=-\partial{G_0}/\partial{\lambda}=
2/(\pi\sqrt{1-\lambda^2})$ and $\partial{G_n}/\partial{\lambda}=
-2\cos(n\theta_0)/(\pi\sqrt{1-\lambda^2})$ ($n\neq 0$). Consequently,
there is always a divergence in the derivatives of the SQC due to
Eq. \eqref{eq2b-10}.

\section{Summary and discussion}\label{sec:4}
To summarize, we have proposed to use the SQC as a signature of QPTs
in the transverse-field \textit{XY} model with three-spin
interaction. The motivation for considering such a quantumness
measure is that it is long ranged and exists in the parameter
regions for which there are no quantum correlations. Compared with
other signatures of QPTs such as entanglement and quantum discord,
our method is powerful due to the following advantages:
(\romannumeral+1) The SQC and its derivative succeed in detecting
precisely all the QPTs in the considered models. (\romannumeral+2)
The effectiveness of SQC in detecting QPTs is independent of the
distance of two spins, which makes it convenient for practical use
as one can choose any two spins other than the restricted
short-distance spins. This also differentiates it from concurrence and
quantum discord, which decrease rapidly with the increasing distance
of two spins and disappear or become infinitesimal when the distance
is long. (\romannumeral+3) The SQC is analytically solvable and
could be estimated experimentally by local projective measurements
and one-qubit tomography. Moreover, the advantage of the SQC method
over the simple coherence method may originate from the fact that
while quantum coherence reveals only the quantum nature of the whole
system under a fixed basis, the SQC takes into account the three
mutually unbiased bases and the local operation and classical
communication between \textit{A} and \textit{B}.
As a consequence, it captures a kind of correlation which contains more
comprehensive information than that of coherence \cite{SQC,naqc2,
naqc3}, hence it is capable of distinguishing the subtle nature of a
system and is more reliable in reflecting the quantum critical
behaviors even when the coherence measures fail to do so.

As the three-spin interaction Hamiltonian may be generated in
optical lattices \cite{three}, we expect our observation can be
confirmed in future experiments with state-of-the-art techniques.
One step further would be to use the SQC method to investigate QPTs
of high-dimensional spin systems and exotic quantum phases in
many-body systems such as topological phase transitions \cite{topo1,
topo2,topo3,topo4,topo5}. Moreover, it is also appealing to study the
dynamics of the SQC, which may provide an interesting scenario for
understanding quantum criticality of many-body systems
\cite{dyqc1,dyqc2,dyqc3}.

\section*{ACKNOWLEDGMENTS}
This work was supported by National Natural Science Foundation of
China (Grant Nos. 11675129, 11774406, and 11934018), National Key R
\& D Program of China (Grant Nos. 2016YFA0302104 and
2016YFA0300600), Strategic Priority Research Program of Chinese
Academy of Sciences (Grant No. XDB28000000), Research Program of
Beijing Academy of Quantum Information Sciences (Grant No. Y18G07),
the New Star Team of XUPT, and the Innovation Fund for graduates
(Grant No. CXJJLA2018007).

\newcommand{\PRL}{Phys. Rev. Lett. }
\newcommand{\RMP}{Rev. Mod. Phys. }
\newcommand{\PRA}{Phys. Rev. A }
\newcommand{\PRB}{Phys. Rev. B }
\newcommand{\PRE}{Phys. Rev. E }
\newcommand{\PRX}{Phys. Rev. X }
\newcommand{\NJP}{New J. Phys. }
\newcommand{\JPA}{J. Phys. A }
\newcommand{\JPB}{J. Phys. B }
\newcommand{\OC}{Opt. Commun.}
\newcommand{\PLA}{Phys. Lett. A }
\newcommand{\EPJB}{Eur. Phys. J. B }
\newcommand{\EPJD}{Eur. Phys. J. D }
\newcommand{\NP}{Nat. Phys. }
\newcommand{\NC}{Nat. Commun. }
\newcommand{\EPL}{Europhys. Lett. }
\newcommand{\AoP}{Ann. Phys. (N.Y.) }
\newcommand{\QIC}{Quantum Inf. Comput. }
\newcommand{\QIP}{Quantum Inf. Process. }
\newcommand{\CPB}{Chin. Phys. B }
\newcommand{\IJTP}{Int. J. Theor. Phys. }
\newcommand{\IJQI}{Int. J. Quantum Inf. }
\newcommand{\IJMPB}{Int. J. Mod. Phys. B }
\newcommand{\PR}{Phys. Rep. }
\newcommand{\SR}{Sci. Rep. }
\newcommand{\LPL}{Laser Phys. Lett. }
\newcommand{\SCG}{Sci. China Ser. G }
\newcommand{\JMP}{J. Math. Phys. }
\newcommand{\RPP}{Rep. Prog. Phys. }
\newcommand{\PA}{Physica A }
%


\begin{thebibliography}{50}

\bibitem{Ficek} Z. Ficek and S. Swain, \emph{Quantum Interference and Coherence: Theory and Experiments}, Springer Series in Optical Sciences (Springer, New York, 2005).
\bibitem{ther5} G. Gour, M. P. M\"{u}ller, V. Narasimhachar, R. W. Spekkens, and N. Y. Halpern, \PR {\bf 583}, 1 (2015).
\bibitem{coher} T. Baumgratz, M. Cramer, and M. B. Plenio, \PRL {\bf 113}, 140401 (2014).
\bibitem{Plenio} A. Streltsov, G. Adesso, and M. B. Plenio, \RMP {\bf 89}, 041003 (2017).
\bibitem{Hu} M. L. Hu, X. Hu, J. C. Wang, Y. Peng, Y. R. Zhang, and H. Fan, \PR {\bf 762-764}, 1 (2018).
\bibitem{qsm} A. Streltsov, E. Chitambar, S. Rana, M. N. Bera, A. Winter, and M. Lewenstein, \PRL {\bf 116}, 240405 (2016).
\bibitem{DQC1} J. Ma, B. Yadin, D. Girolami, V. Vedral, and M. Gu, \PRL {\bf 116}, 160407 (2016).
\bibitem{DJ} M. Hillery, \PRA {\bf 93}, 012111 (2016).
\bibitem{Grover} H. L. Shi, S. Y. Liu, X. H. Wang, W. L. Yang, Z. Y. Yang, and H. Fan, \PRA {\bf 95}, 032307 (2017).
\bibitem{path1} M. N. Bera, T. Qureshi, M. A. Siddiqui, and A. K. Pati, \PRA {\bf 92}, 012118 (2015).

\bibitem{path2} E. Bagan, J. A. Bergou, S. S. Cottrell, and M. Hillery, \PRL {\bf 116}, 160406 (2016).
\bibitem{coher-ent} A. Streltsov, U. Singh, H. S. Dhar, M. N. Bera, and G. Adesso, \PRL {\bf 115}, 020403 (2015).
\bibitem{convex3} X. Qi, T. Gao, and F. Yan, \JPA {\bf 50}, 285301 (2017).
\bibitem{SQC} D. Mondal, T. Pramanik, and A. K. Pati, \PRA {\bf 95}, 010301(R) (2017).
\bibitem{naqc2} M. L. Hu and H. Fan, \PRA {\bf 98}, 022312 (2018).
\bibitem{naqc3} M. L. Hu, X. M. Wang, and H. Fan, \PRA {\bf 98}, 032317 (2018).
\bibitem{Tan} K. C. Tan, H. Kwon, C. Y. Park, and H. Jeong, \PRA {\bf 94}, 022329 (2016).
\bibitem{Yao} Y. Yao, X. Xiao, L. Ge, and C. P. Sun, \PRA {\bf 92}, 022112 (2015).
\bibitem{Hufan} M. L. Hu and H. Fan, \PRA {\bf 95}, 052106 (2017).
\bibitem{Hux1} X. Hu, A. Milne, B. Zhang, and H. Fan, \SR {\bf 6}, 19365 (2015).

\bibitem{Yuc} J. Zhang, S. R. Yang, Y. Zhang, and C. S. Yu, \SR {\bf 7}, 45598 (2017)
\bibitem{Hux2} X. Hu and H. Fan, \SR {\bf 6}, 34380 (2016).
\bibitem{chenj} J. J. Chen, J. Cui, Y. R. Zhang, and H. Fan, \PRA {\bf 94}, 022112 (2016).
\bibitem{Qin} M. Qin, Z. Ren, and X. Zhang, \PRA {\bf 98}, 012303 (2018).
\bibitem{spin1} A. L. Malvezzi, G. Karpat, B. C. \c{C}akmak, F. F. Fanchini, T. Debarba, and R. O. Vianna, \PRB {\bf 93}, 184428 (2016).
\bibitem{skif} D. Girolami, \PRL {\bf 113}, 170401 (2014).
\bibitem{Dubai} S. Du and Z. Bai, \AoP {\bf 359}, 136 (2015).
\bibitem{Karpat} G. Karpat, B. \c{C}akmak, and F. F. Fanchini, \PRB {\bf 90}, 104431 (2014).
\bibitem{Leisg} S. G. Lei and P. Q. Tong, \QIP {\bf 15}, 1811 (2016).
\bibitem{Liyc} Y. C. Li and H. Q. Lin, \SR {\bf 6}, 26365 (2016).

\bibitem{Ywl} T. C. Yi, W. L. You, N. Wu, and A. M. Ole\'{s}, \PRB {\bf 100}, 024423 (2019).
\bibitem{EoF} W. K. Wootters, \PRL {\bf 80}, 2245 (1998).
\bibitem{nature} A. Osterloh, L. Amico, G. Falci, and R. Fazio, Nature (Londan) {\bf 416}, 608 (2002).
\bibitem{Osborne} T. J. Osborne and M. A. Nielsen, \PRA {\bf 66}, 032110 (2002).
\bibitem{Gusj1} S. J. Gu, H. Q. Lin, and Y. Q. Li, \PRA {\bf 68}, 042330 (2003).
\bibitem{Gusj2} S. J. Gu, G. S. Tian, and H. Q. Lin, \PRA {\bf 71}, 052322 (2005).
\bibitem{Amico} L. Amico, R. Fazio, A. Osterloh, and V. Vedral, \RMP {\bf 80}, 517 (2008).
\bibitem{QD} H. Ollivier and W. H. Zurek, \PRL {\bf 88}, 017901 (2001).
\bibitem{QD2} L. Henderson and V. Vedral, \JPA {\bf 34}, 6899 (2001).
\bibitem{XXZ} T. Werlang, C. Trippe, G. A. P. Ribeiro, and G. Rigolin, \PRL {\bf 105}, 095702 (2010).

\bibitem{Sarandy} M. S. Sarandy, \PRA {\bf 80}, 022108 (2009).
\bibitem{Ising} R. Dillenschneider, \PRB {\bf 78}, 224413 (2008).
\bibitem{XY} J. Maziero, H. C. Guzman, L. C. C\'{e}leri, M. S. Sarandy, and R. M. Serra, \PRA {\bf 82}, 012106 (2010).
\bibitem{XYthree} Y. C. Li and H. Q. Lin, \PRA {\bf 83}, 052323 (2011).
\bibitem{XYDM} B. Q. Liu, B. Shao, J. G. Li, J. Zou, and L. A. Wu, \PRA {\bf 83}, 052112 (2011).
\bibitem{qd-np} Y. Huang, \NJP {\bf 16}, 033027 (2014).
\bibitem{qdtwo} D. Girolami and G. Adesso, \PRA {\bf 83}, 052108 (2011).
\bibitem{qcexp1} Y. T. Wang, J. S. Tang, Z. Y. Wei, S. Yu, Z. J. Ke, X. Y. Xu, C. F. Li, and G. C. Guo, \PRL {\bf 118}, 020403 (2017).
\bibitem{qcexp2} D. J. Zhang, C. L. Liu, X. D. Yu, and D. M. Tong, \PRL {\bf 120}, 170501 (2018).
\bibitem{qcexp3} X. D. Yu and O. G\"{u}hne, \PRA {\bf 99}, 062310 (2019).

\bibitem{three} J. K. Pachos and M. B. Plenio, \PRL {\bf 93}, 056402 (2004).
\bibitem{QPTs2} S. Sachdev, \emph{Quantum Phase Transitions} (Cambridge University Press, Cambridge, England, 2000).
\bibitem{epjb} I. Titvinidze and G. I. Japaridze, \EPJB {\bf 32}, 383 (2003).
\bibitem{Wang} X. G. Wang, \PLA {\bf 331}, 164 (2004).
\bibitem{Gusj} S. J. Gu, C. P. Sun, and H. Q. Lin, \JPA {\bf 41}, 025002 (2008).
\bibitem{magnet} E. Barouch, B. M. McCoy, and M. Dresden, \PRA {\bf 2}, 1075 (1970).
\bibitem{xyt1} E. Barouch and B. McCoy, \PRA {\bf 3}, 786 (1971).
\bibitem{phase} P. Pfeuty, \AoP {\bf 57}, 79 (1970).
\bibitem{QPTxy} M. Zhong and P. Tong, \JPA {\bf 43}, 505302 (2010).
\bibitem{xyt2} B. McCoy, E. Barouch, and D. Abraham, \PRA {\bf 4}, 2331 (1971).

\bibitem{topo1} A. Kitaev and J. Preskill, \PRL {\bf 96}, 110404 (2006).
\bibitem{topo2} A. Hamma, W. Zhang, S. Haas, and D. A. Lidar, \PRB {\bf 77}, 155111 (2008).
\bibitem{topo3} F. Pollmann, A. M. Turner, E. Berg, and M. Oshikawa, \PRB {\bf 81}, 064439 (2010).
\bibitem{topo4} Y. X. Chen and S. W. Li, \PRA {\bf 81}, 032120 (2010).
\bibitem{topo5} J. Cui, J. P. Cao, and H. Fan, \PRA {\bf 82}, 022319 (2010).
\bibitem{dyqc1} H. T. Quan, Z. Song, X. F. Liu, P. Zanardi, and C. P. Sun, \PRL {\bf 96}, 140604 (2006).
\bibitem{dyqc2} D. Rossini, T. Calarco, V. Giovannetti, S. Montangero, and R. Fazio, \PRA {\bf 75}, 032333 (2007).
\bibitem{dyqc3} Z. Sun, X. G. Wang, and C. P. Sun, \PRA {\bf 75}, 062312 (2007).

\end{thebibliography}

\end{document}